\documentclass[aps,pra,reprint,showpacs,groupedaddress]{revtex4-1}
\usepackage{graphicx}
\usepackage{hyperref}

\begin{document}

\title{Realization of the purely spatial Einstein-Podolsky-Rosen paradox in full-field images of spontaneous parametric down conversion.}

\author{Paul-Antoine Moreau}
\author{Jo\'{e} Mougin-Sisini}
\author{Fabrice Devaux}
\author{Eric Lantz}
\affiliation{Institut FEMTO-ST  D\'epartement d'Optique PM Duffieux
\\UMR CNRS - Universit\'e de Franche-Comt\'e $n\,^{\circ}\mathrm{ 6174}$,
Route de Gray  25030 Besan\c con Cedex, FRANCE}

\date{\today}

\begin{abstract}
We demonstrate Einstein-Podolsky-Rosen (EPR) entanglement by detecting purely spatial quantum correla-tions in the near and far fields of spontaneous parametric down-conversion generated in a type-2 beta barium borate crystal. Full-field imaging is performed in the photon-counting regime with an electron-multiplying CCD camera. The data are used without any postselection, and we obtain a violation of Heisenberg inequalities with inferred quantities taking into account all the biphoton pairs in both the near and far fields by integration on the entire two-dimensional transverse planes. This ensures a rigorous demonstration of the EPR paradox in its original position-momentum form.
\end{abstract}


\maketitle

\section{Introduction} In 1935, Einstein, Podolsky, and Rosen (EPR) proposed a \textit{Gedankenexperiment} \cite{Einstein1935} involving two spatially
separated but entangled particles. They showed that
quantum mechanics predicts that these particles could
have both perfectly correlated positions and momenta, in
contradiction with the so-called \textit{local realism} where two
distant particles should be treated as two different sys-
tems. Though the original intention of EPR was to show
that quantum mechanics is not complete, the standard
present view is that entangled particles do experience
nonlocal correlations \cite{Bell1964,Aspect1981}.

Recently, the realization and detection of entangled EPR states aroused much interest of the scientific community fol-lowing a testable formulation of the EPR paradox introduced by Reid \cite{Reid1989}, involving correlations of quadratures of twin
beams. A recent review on the subject has been given in Ref. \cite{Reid2009}. Furthermore, position and momentum entanglement of beams have been demonstrated at the EPR level by combin-ing squeezed light from two spatial modes, with measurements by homodyne detection in the temporal domain \cite{Wagner2008,Boyer2008}.

Spontaneous parametric down-conversion (SPDC) pro-vides independent pairs of entangled photons that makes the systemvery close to that considered in the original EPR paper: the positions of photons 1 and 2 are detected in the near field,
and their momenta correspond to the far field. Howell\textit{et al.} \cite{Howell2004} havemeasured in both planes the probability distribution of the
position of photon 2, conditioned by the detection of photon
1. The product of the conditional variances was found to be
25 times smaller than the limit for the product of variances for
a single photon given by Heisenberg's uncertainty relation.
This impressive result was obtained by measuring tempo-ral coincidences between cross-polarized photons in type-2
SPDC. These photons were separated by a polarizing beam
splitter: for a fixed position of a narrow slit transmitting
photon 1 to an avalanche photodiode, the level of coincidences
was measured for each position of a similar slit transmitting
photon 2 to a separate similar detector. Hence, in the words
of Reid \textit{et al.} \cite{Reid2009}, "detection events are only considered if
two emitted photons are simultaneously detected". In this
sense, they did not prospect the full EPR characteristic of
SPDC but a monodimensional and point EPR paradox using postselected data. The same comment could be made on the
recent realization by Leach \textit{et al.} \cite{Leach2012}, where point detection
was used in one of the transverse dimensions of the field.

We present here a full spatial demonstrationof EPRsteering
by imaging highly spatially multimode type-2 SPDC with an
electron-multiplying CCD (EMCCD) camera. This ensures
taking into account each detection event that occurs in a
relatively long exposure time compared to the laser pulse
duration and, more importantly, compared to the coincidence
time detection used in the experiments that use correlation data
detection. Hence, even if there are losses and false-positive and
-negative events, these spurious events are random, which is
fundamentallydifferent fromconsideringapriorithat pairs are
correlated and detecting only in the temporal and spatial gate
where the twinphoton arrives. Second,Heisenberg inequalities
concern a sole system, hence it is meaningless to test these
inequalities if the near field and the far field do not correspond
to this unique system. By treating full-field bidimensional
images of photodetection and measuring variances in two
orthogonal directions, we assure a perfect correspondence
between the subsystems involved in the near field and in
the far field \cite{Reid2009}, in contrast to point or one-dimensional (1D)
detection. Moreover we demonstrate not only an EPR paradox
in two spatial transverse dimensions but also one using an
isotropic criterion involving the whole system. Note that the
use of type-2 phase matching allows us to spatially separate
the idler and signal photons to be close to the real conditions
of an EPR test of local realism. Detection is performed
with an EMCCD camera. The ability of EMCCD cameras
to reach the photon-counting regime with a very high quantum
efficiency \cite{Lantz2008,Jedrkiewicz2012} makes them useful in quantum optics, and
they have already been used by our group to characterize
quantum correlations in the far field of type-1 \cite{Blanchet2008,Blanchet2010} and  type-2 \cite{Devaux2012} SPDC. Note that previous work has been done
by other groups to detect the quantum correlation in the far
field \cite{Jost1998,Oemrawsingh2002}, using intensified CCD (ICCD) cameras that have
a lower quantum efficiency.

\section{Theory}
SPDC induced by a wide monomode Gaussian pump
is a strongly multimode beam: the extension of the down-converted beam in the near field (image plane) is
identical to that of the pump in the limit of low gain and for a
sufficiently wide and thin crystal, while the far-field (Fourier
plane) extension is limited by phase matching. The etendue
of the beam, i.e., the product of its transverse surface by the
solid angle it subtends or the number of transverse modes
in appropriate units, corresponds to the two-photon Schmidt
number \cite{Exter2006}.

The spatial extension of a mode in either the near or
the far field is proportional to the inverse of the full beam
extension in the other plane. For single-photon imaging, the
laws of diffraction are equivalent to Heisenberg's uncertainty
relation: a photon that can be localized in one mode of the
near field, for example, by traversing an aperture of the size
corresponding to the mode, will be detected at a random
position in the entire far-field diffraction pattern. However,
the laws of quantum mechanics state that a pair of signal-idler
photons will be detected either in the same mode in the near
field or in opposite modes in the far field if no detection
occurs in the other plane. Because the detection plane can
be chosen at a time when causal interaction between photons
is no longer possible, these correlations are not compatible
with local realism, as demonstrated first in the EPR paper \cite{Einstein1935}, though compatible with Heisenberg's uncertainty relation
since correlations cannot be measured in both planes for the
same photon pair.

We can describe the SPDC behavior as follows: for a detection of a photon 1 at $\mathbf{r_{1}}$, the probability density of detection of a photon 2 at
$\mathbf{r_{2}}$ can be written as:
\begin{equation}\label{p1}
p\left(\mathbf{r_{2}}\right|\left.\mathbf{r_{1}}\right)=p\left(\mathbf{r_{2}}\right)+ f\left(\mathbf{\Delta r}\right)
\end{equation}
where $p(\mathbf{r_{2}})$ is the probability density of
detecting a photon of another pair (accidental coincidences)
and $f(\mathbf{\Delta r})$ is the probability density of detection of the
twin photon, with $\mathbf{\Delta r}
=\|\mathbf{r_{2}}\pm\mathbf{r_{1}}\|$, plus sign indicating the far field (correlation of
momenta on oppositemodes) and theminus sign the near field.
Translational invariance, circular symmetry, and independence
of the pairs (pure SPDC without further amplification) are
assumed. Hence, if N$_1$ is the number of photons 1 detected on a surface S$_1$ and N$_2$ the corresponding quantity for photons 2, we have:
\begin{eqnarray}\label{n1n2}
\left<N_1 N_2\right>&=&\int_{S_1}{d^{2}\mathbf{r_{1}}\int_{S_2}{d^{2}\mathbf{r_{2}}\
p\left(\mathbf{r_{1}},\mathbf{r_{2}}\right)}}\\
&=&\int_{S_1}{d^{2}\mathbf{r_{1}}\int_{S_2}{d^{2}\mathbf{r_{2}}\left\{
p\left(\mathbf{r_{1}}\right) p\left(\mathbf{r_{2}}\right)+
p\left(\mathbf{r_{1}}\right)f\left(\mathbf{\Delta r}\right)\right\}}}\nonumber
\end{eqnarray}
Therefore, the probability of detection in S$_2$ of the twin photon 2 of the photon 1 detected on S$_1$ is simply given by:
\begin{equation}\label{fdeltar}
F(S_2)=\int_{S_2}d^{2}\mathbf{r_{2}} \ f(\mathbf{\Delta r})=\frac{\left<N_1 N_2\right>-\left<N_1\right>\left<N_2\right>}{\left<N_1\right>}
\end{equation}
If S$_1$ and S$_2$ are the same size, this expression can be
symmetrized and becomes the normalized intercorrelation function:
\begin{equation}\label{fs2}
F(S_2)=F(S_1)=\frac{\left<N_1 N_2\right>-\left<N_1\right>\left<N_2\right>}{\frac{1}{2}(\left<N_1\right>+\left<N_2\right>)}
\end{equation}
Themean in this equation can be estimated by spatial averages
on the different pixels of the image for a fixed $\mathbf{\Delta r}$, given by the intercorrelation of two "regions of interest" (ROIs) of an image, each one corresponding to one polarization of
the SPDC. We will therefore obtain a nonlocal estimation
involving all the light. Note that deterministic spatial variations
of the mean intensity do not preclude the validity of these
spatial averages, inasmuch as the width of the intercorrelation
function is smaller than the scale of this deterministic variation.
Indeed, the covariance signal-idler for a region formed by
independent area is the sum of the covariances of each area,
just as the mean for the region is the sumof themeans for each
area. Hence, if the ratio between the covariance and the mean
intensity does not depend of this mean, it will be retrieved by
spatial averaging even if the mean varies spatially. Because
of the weak signal-to-noise ratio, we proceed to an additional
statistical average on different images taken at different times
for the same system configuration.\\
For independent pairs, the quantity in equation (\ref{fs2}) can be expressed as a function of the variance of the difference between N$_1$ and N$_2$:
\begin{eqnarray}\label{fdeltarvmoins}
\left<N_1\right>=\left<N_2\right>=\left<N_1^2\right>-\left<N_1\right>^2\nonumber\\
\Rightarrow\ F(S_2)=1-\frac{\left<(N_1- N_2)^2\right>}{\left<N_1+N_2\right>}
\end{eqnarray}

The physical quantities used to test the EPR violation of Heisenberg inequalities are the spatial variances in each dimension $\Delta^2x$ and $\Delta^2y$. To define a unique global criterion,  we use the mean value of the two variances (i.e. half of the mean of the squared distance) :
\begin{equation}
\Delta^2 r,p=\frac{\Delta^2x,p_x+\Delta^2y,p_y}{2}
\label{tr}
\end{equation}

Thus, by introducing the Heisenberg inequalities in the product
of near-field and far-field variances, one gets
\begin{equation}
\Delta^2r\Delta^2p=\frac{1}{4}(\Delta^2x+\Delta^2y)(\Delta^2p_x+\Delta^2p_y)\geq\frac{\hbar^2}{4}
\label{heis2d}
\end{equation}
We have used the fact that $\Delta^2x\Delta^2p_y+\Delta^2y\Delta^2p_x\geq2\frac{\hbar^2}{4}$. Note that this two dimensional diffraction limit will only be reached by isotropic 2-D gaussians.

\section{Experiment}
\begin{figure}
\resizebox{1\columnwidth}{!}{\includegraphics{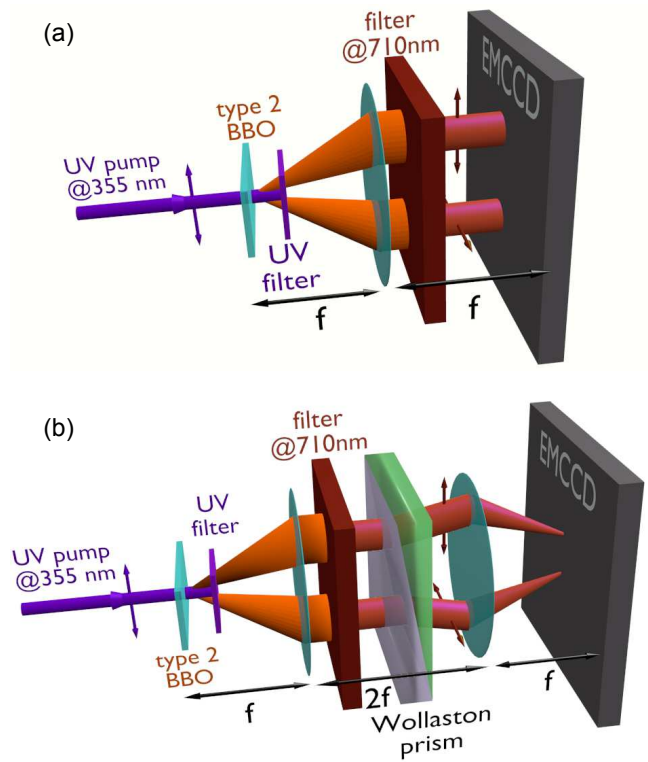}}
\caption{Experimental setups. (a) Far field. (b) Near field. Brown arrows give the polarization directions}
\label{setup}
\end{figure}

The experimental setup is represented in Fig. \ref{setup}. The pump pulse at 355 nm provided by a passively Q-switched Nd:YAG laser (mean power: 27 mW, pulse duration: 300 ps, repetition rate: 1 kHz), illuminates a 1-mm long type 2 BBO nonlinear crystal. The far-field image of the SPDC is formed on the EMCCD in the focal plane of a $37\ mm$ focal aspheric lens: Fig. \ref{setup}(a). In the near field configuration, Fig \ref{setup}(b), the signal and idler photons are separated by a Wollaston prism of $1.5°$ of angular separation positioned around the Fourier plane.  The plane in the middle of the BBO crystal is imaged on the EMCCD plane by  a second identical aspheric lens, in order to minimize the walk-off effects \cite{Brambilla2004}.  The transversal magnification has been checked:  $\gamma=1.003\pm0.005$.  The back-illuminated EMCCD camera from Andor Technology (model iXon+ DU897-ECS-BV) has a quantum efficiency greater than 90\% in the visible range. The detector area is formed by 512$\times$512 pixels, with a pixel size of $s_{pix}=16\times16 \mu$m$^2$ (i.e. 0.46$\times$0.46 mrad$^2$ in the far field after division by the focal length). We used a readout rate of 10 MHz at 14 bits and the camera was cooled to -85$^\circ$C. Measurements were performed for a crystal orientation corresponding to non critical phase matching  at degeneracy, i.e. collinear orientation of the signal and idler Poynting vectors in the crystal \cite{Lantz2000}. Photon pairs emitted around the degeneracy are selected by mean of a narrow-band interferential filter centered at 710 nm ($\Delta\lambda$=4 nm).
As in \cite{Blanchet2010} and \cite{Devaux2012}, the photon-counting regime is ensured by adjusting the exposure time in such a way that the mean fluence of SPDC was between 0.1 and 0.2 photon per pixel, in order to minimize the false detections \cite{Lantz2008}. Moreover, the use of pump pulses with 300 ps duration (much longer than the coherence time of SPDC) and an exposure time of the EMCCD of 10 ms (i.e. 10 laser shots) allow the excess noise to be limited by increasing the number of temporal modes \cite{Brambilla2008} : the mean number of photons for one spatiotemporal mode is less than $10^{-3}$, in good agreement with the hypothesis of pure spontaneous parametric down conversion, without any stimulated amplification.

\subsection{Near field}

\begin{figure}
\resizebox{1\columnwidth}{!}{\includegraphics{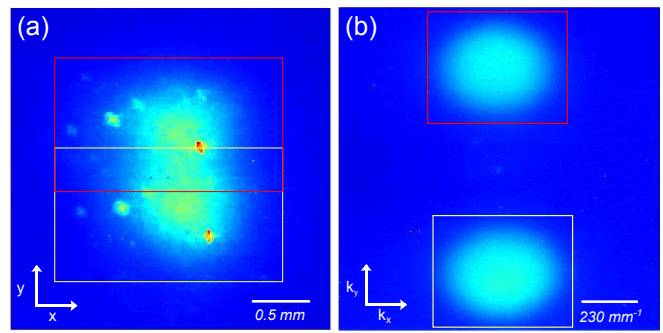}}
\caption{Intensity distributions. (a) Near field. (b) Far field. In both figures rectangles denote the ROIs used for calculation.}
\label{Int}
\end{figure}

\begin{figure}
\resizebox{1\columnwidth}{!}{\includegraphics{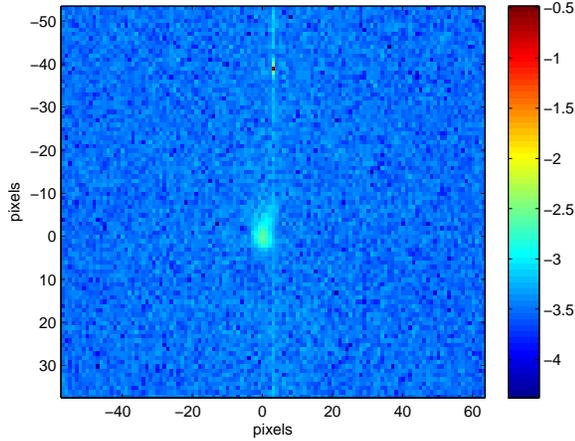}}
\caption{Near-field complete intercorrelation in log scale.}
\label{Cpauto}
\end{figure}

\begin{figure}
\resizebox{1\columnwidth}{!}{\includegraphics{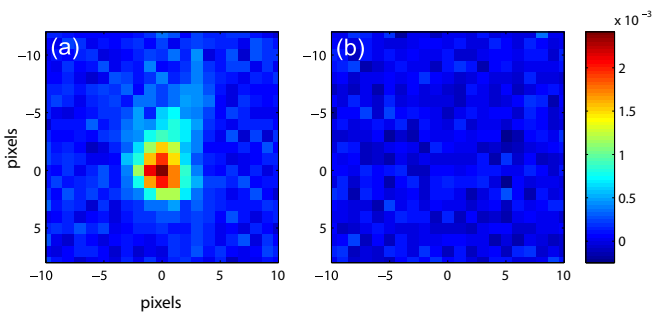}}
\caption{(a) Near-field intercorrelation function. (b) Corresponding witness intercorrelation.}
\label{Cp}
\end{figure}

We show in Fig. \ref{Int} (a) the sum of the 10000 near field images.  The two SPDC patterns are clearly visible, with inhomogeneities and hot spots due to defaults on the crystal . The ROIs corresponding to either the signal or the idler, large enough  to encompass all the light for each polarization,  have a common area. As a consequence, the intercorrelation function exhibits a strong autocorrelation peak, as can be seen in Fig. \ref{Cpauto}, but the intercorrelation peak, due to quantum correlations, and the autocorrelation peak  are clearly distinguishable. We use  a Fourier algorithm without any zero padding to compute the intercorrelation, which is equivalent to a periodisation of the images.
Figure \ref{Cp} presents the near-field intercorrelation peak ob-tained from ROIs taken in the same image and a witness
intercorrelation obtained from ROIs taken in two successive
images. The absence of any non-negligible intercorrelation
value in the second image shows that the inhomogeneities in
the crystal do not create any deterministic intercorrelation pat-tern. The first image exhibits a weak intercorrelation vertical
line (more visible in \ref{Cpauto} because of the log scale), which
is due to the spreading of the autocorrelation peak induced
by smearing in the gain register of the EMCCD camera. The
existence of such an artifact again shows the importance of
sufficiently separating the signal and idler photons on the
camera to avoid any superposition of the autocorrelation and
intercorrelation peaks.
By fitting the intercorrelation peakwith
a two-dimensional Gaussian function, we estimate the inferred
near-field standard deviations in pixels:
\begin{equation}
\Delta x=1.53\pm0.07,\ \ \Delta y=2.2\pm0.1,\ \ \Delta r=1.89\pm0.09
\end{equation}
By integrating thefitted curve,we alsoobtain the total quantum
correlation coefficient in the near field,
 $R_{n}=5\times10^{-2}$. This
coefficient corresponds to the intercorrelation of pixels much
larger than the coherence area \cite{Devaux2012}.

\subsection{Far field}
\begin{figure}
\resizebox{1\columnwidth}{!}{\includegraphics{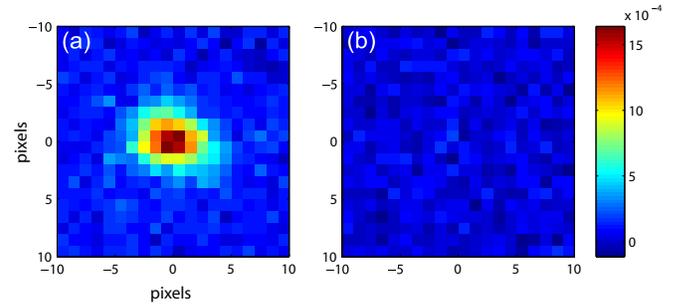}}
\caption{(a) Far-field intercorrelation function. (b) Corresponding witness intercorrelation.}
\label{Cl}
\end{figure}

Figure \ref{Int} (b) shows the sum of 10000 images in the far-field
configuration. The intercorrelation function obtained in the far
field is presented in Fig. \ref{Cl}. Note that the anisotropy of the
peak is mainly due to the anisotropy of the shape of the pump.
However, as predicted by simulations, an enlargement exists
due to the imperfect degeneracy of the photons wavelength.
This enlargement is itself anisotropic and, as predicted by the
theory, is greater in the walk-off direction which separates the
two fluorescence spots in the far field (vertical direction on
each image presented here). The experimental results are in
agreement with this phenomenon since the enlargement of the
intercorrelation peak is greater in the vertical dimension than
in the horizontal one \cite{Devaux2012}.
We finally find the inferred standard
deviations in far-field pixel units:
\begin{equation}
\Delta p_x=2.35\pm0.08,\ \ \Delta p_y=1.85\pm0.07,\ \ \Delta p=2.11\pm0.07
\end{equation}
The total quantum correlation coefficient in the far field
is $R_{p}=4.4\times10^{-2}$. In agreement with Eq. \ref{fdeltarvmoins}, we have experimentally shown in  \cite{Devaux2012} that $(1-R_{p})$ is equal to the variance of the difference between areas greater than the coherence cell, expressed in shot noise units.

\section{EPR violation of the Heisenberg inequalities}

We are now able to test the violation of the Heisenberg
inequalities by using the inferred quantitieswe havemeasured. In the horizontal direction, we have:
\begin{eqnarray}
\Delta^2x\Delta^2p_x&=&\left(1.53\times2.35\frac{2\pi\cdot s_{pix}}{f\cdot\lambda}\hbar\right)^2\nonumber\\
&=&(0.048\pm0.008)\hbar^2<\frac{\hbar^2}{4}
\end{eqnarray}
giving a violation factor of $5.2\pm0.8$.

In the vertical direction, we find:
\begin{equation}
\Delta^2y\Delta^2p_y=(0.06\pm0.01)\hbar^2<\frac{\hbar^2}{4}
\end{equation}
giving  a violation factor of $4\pm1$.

And finally, by using the complete statistic of the fluorescence, from (\ref{tr}) and (\ref{heis2d}) one obtain:
\begin{equation}
\Delta^2r\Delta^2p=(0.06\pm0.01)\hbar^2<\frac{\hbar^2}{4}
\end{equation}
which gives a violation factor of $4\pm1$.
\section{Conclusion}
We have demonstrated a purely spatial EPR paradox by
using a full-field and direct detection method. The violation
of the Heisenberg inequalities for the inferred quantities has
been demonstrated for the whole system involving the two
spatial transverse dimensions. By recording all the photon
pairs generated by spontaneous parametric down-conversion
in the near field and the far field of the same system, we
make fewer supplementary assumptions than previous papers
that involve postselection or homodyne detection. Hence, this
demonstration of one of the most fascinating phenomena of
quantum mechanics is made in the form closest to its original
formulation.

\begin{acknowledgments}
Some of the computations have been performed on the supercomputer facilities of the M\'{e}socentre
de calcul de Franche-Comt\'{e}.
\end{acknowledgments}

\begin{thebibliography}{20}%
\makeatletter
\providecommand \@ifxundefined [1]{%
 \@ifx{#1\undefined}
}%
\providecommand \@ifnum [1]{%
 \ifnum #1\expandafter \@firstoftwo
 \else \expandafter \@secondoftwo
 \fi
}%
\providecommand \@ifx [1]{%
 \ifx #1\expandafter \@firstoftwo
 \else \expandafter \@secondoftwo
 \fi
}%
\providecommand \natexlab [1]{#1}%
\providecommand \enquote  [1]{``#1''}%
\providecommand \bibnamefont  [1]{#1}%
\providecommand \bibfnamefont [1]{#1}%
\providecommand \citenamefont [1]{#1}%
\providecommand \href@noop [0]{\@secondoftwo}%
\providecommand \href [0]{\begingroup \@sanitize@url \@href}%
\providecommand \@href[1]{\@@startlink{#1}\@@href}%
\providecommand \@@href[1]{\endgroup#1\@@endlink}%
\providecommand \@sanitize@url [0]{\catcode `\\12\catcode `\$12\catcode
  `\&12\catcode `\#12\catcode `\^12\catcode `\_12\catcode `\%12\relax}%
\providecommand \@@startlink[1]{}%
\providecommand \@@endlink[0]{}%
\providecommand \url  [0]{\begingroup\@sanitize@url \@url }%
\providecommand \@url [1]{\endgroup\@href {#1}{\urlprefix }}%
\providecommand \urlprefix  [0]{URL }%
\providecommand \Eprint [0]{\href }%
\providecommand \doibase [0]{http://dx.doi.org/}%
\providecommand \selectlanguage [0]{\@gobble}%
\providecommand \bibinfo  [0]{\@secondoftwo}%
\providecommand \bibfield  [0]{\@secondoftwo}%
\providecommand \translation [1]{[#1]}%
\providecommand \BibitemOpen [0]{}%
\providecommand \bibitemStop [0]{}%
\providecommand \bibitemNoStop [0]{.\EOS\space}%
\providecommand \EOS [0]{\spacefactor3000\relax}%
\providecommand \BibitemShut  [1]{\csname bibitem#1\endcsname}%
\let\auto@bib@innerbib\@empty
\bibitem [{\citenamefont {Einstein}\ \emph {et~al.}(1935)\citenamefont
  {Einstein}, \citenamefont {Podolsky},\ and\ \citenamefont
  {Rosen}}]{Einstein1935}%
  \BibitemOpen
  \bibfield  {author} {\bibinfo {author} {\bibfnamefont {A.}~\bibnamefont
  {Einstein}}, \bibinfo {author} {\bibfnamefont {B.}~\bibnamefont {Podolsky}},
  \ and\ \bibinfo {author} {\bibfnamefont {N.}~\bibnamefont {Rosen}},\ }\href
  {\doibase 10.1103/PhysRev.47.777} {\bibfield  {journal} {\bibinfo  {journal}
  {Phys. Rev.}\ }\textbf {\bibinfo {volume} {47}},\ \bibinfo {pages} {777}
  (\bibinfo {year} {1935})}\BibitemShut {NoStop}%
\bibitem [{\citenamefont {Bell}(1964)}]{Bell1964}%
  \BibitemOpen
  \bibfield  {author} {\bibinfo {author} {\bibfnamefont {J.~S.}\ \bibnamefont
  {Bell}},\ }\href@noop {} {\bibfield  {journal} {\bibinfo  {journal}
  {Physics}\ }\textbf {\bibinfo {volume} {1}},\ \bibinfo {pages} {195}
  (\bibinfo {year} {1964})}\BibitemShut {NoStop}%
\bibitem [{\citenamefont {Aspect}\ \emph {et~al.}(1981)\citenamefont {Aspect},
  \citenamefont {Grangier},\ and\ \citenamefont {Roger}}]{Aspect1981}%
  \BibitemOpen
  \bibfield  {author} {\bibinfo {author} {\bibfnamefont {A.}~\bibnamefont
  {Aspect}}, \bibinfo {author} {\bibfnamefont {P.}~\bibnamefont {Grangier}}, \
  and\ \bibinfo {author} {\bibfnamefont {G.}~\bibnamefont {Roger}},\ }\href
  {\doibase 10.1103/PhysRevLett.47.460} {\bibfield  {journal} {\bibinfo
  {journal} {Phys. Rev. Lett.}\ }\textbf {\bibinfo {volume} {47}},\ \bibinfo
  {pages} {460} (\bibinfo {year} {1981})}\BibitemShut {NoStop}%
\bibitem [{\citenamefont {Reid}(1989)}]{Reid1989}%
  \BibitemOpen
  \bibfield  {author} {\bibinfo {author} {\bibfnamefont {M.~D.}\ \bibnamefont
  {Reid}},\ }\href {\doibase 10.1103/PhysRevA.40.913} {\bibfield  {journal}
  {\bibinfo  {journal} {Phys. Rev. A}\ }\textbf {\bibinfo {volume} {40}},\
  \bibinfo {pages} {913} (\bibinfo {year} {1989})}\BibitemShut {NoStop}%
\bibitem [{\citenamefont {Reid}\ \emph {et~al.}(2009)\citenamefont {Reid},
  \citenamefont {Drummond}, \citenamefont {Bowen}, \citenamefont {Cavalcanti},
  \citenamefont {Lam}, \citenamefont {Bachor}, \citenamefont {Andersen},\ and\
  \citenamefont {Leuchs}}]{Reid2009}%
  \BibitemOpen
  \bibfield  {author} {\bibinfo {author} {\bibfnamefont {M.~D.}\ \bibnamefont
  {Reid}}, \bibinfo {author} {\bibfnamefont {P.~D.}\ \bibnamefont {Drummond}},
  \bibinfo {author} {\bibfnamefont {W.~P.}\ \bibnamefont {Bowen}}, \bibinfo
  {author} {\bibfnamefont {E.~G.}\ \bibnamefont {Cavalcanti}}, \bibinfo
  {author} {\bibfnamefont {P.~K.}\ \bibnamefont {Lam}}, \bibinfo {author}
  {\bibfnamefont {H.~A.}\ \bibnamefont {Bachor}}, \bibinfo {author}
  {\bibfnamefont {U.~L.}\ \bibnamefont {Andersen}}, \ and\ \bibinfo {author}
  {\bibfnamefont {G.}~\bibnamefont {Leuchs}},\ }\href {\doibase
  10.1103/RevModPhys.81.1727} {\bibfield  {journal} {\bibinfo  {journal} {Rev.
  Mod. Phys.}\ }\textbf {\bibinfo {volume} {81}},\ \bibinfo {pages} {1727}
  (\bibinfo {year} {2009})}\BibitemShut {NoStop}%
\bibitem [{\citenamefont {Wagner}\ \emph {et~al.}(2008)\citenamefont {Wagner},
  \citenamefont {Janousek}, \citenamefont {Delaubert}, \citenamefont {Zou},
  \citenamefont {Harb}, \citenamefont {Treps}, \citenamefont {Morizur},
  \citenamefont {Lam},\ and\ \citenamefont {Bachor}}]{Wagner2008}%
  \BibitemOpen
  \bibfield  {author} {\bibinfo {author} {\bibfnamefont {K.}~\bibnamefont
  {Wagner}}, \bibinfo {author} {\bibfnamefont {J.}~\bibnamefont {Janousek}},
  \bibinfo {author} {\bibfnamefont {V.}~\bibnamefont {Delaubert}}, \bibinfo
  {author} {\bibfnamefont {H.}~\bibnamefont {Zou}}, \bibinfo {author}
  {\bibfnamefont {C.}~\bibnamefont {Harb}}, \bibinfo {author} {\bibfnamefont
  {N.}~\bibnamefont {Treps}}, \bibinfo {author} {\bibfnamefont {J.~F.}\
  \bibnamefont {Morizur}}, \bibinfo {author} {\bibfnamefont {P.~K.}\
  \bibnamefont {Lam}}, \ and\ \bibinfo {author} {\bibfnamefont {H.~A.}\
  \bibnamefont {Bachor}},\ }\href {\doibase 10.1126/science.1159663} {\bibfield
   {journal} {\bibinfo  {journal} {Science}\ }\textbf {\bibinfo {volume}
  {321}},\ \bibinfo {pages} {541} (\bibinfo {year} {2008})}\BibitemShut
  {NoStop}%
\bibitem [{\citenamefont {Boyer}\ \emph {et~al.}(2008)\citenamefont {Boyer},
  \citenamefont {Marino}, \citenamefont {Pooser},\ and\ \citenamefont
  {Lett}}]{Boyer2008}%
  \BibitemOpen
  \bibfield  {author} {\bibinfo {author} {\bibfnamefont {V.}~\bibnamefont
  {Boyer}}, \bibinfo {author} {\bibfnamefont {A.~M.}\ \bibnamefont {Marino}},
  \bibinfo {author} {\bibfnamefont {R.~C.}\ \bibnamefont {Pooser}}, \ and\
  \bibinfo {author} {\bibfnamefont {P.~D.}\ \bibnamefont {Lett}},\ }\href
  {\doibase 10.1126/science.1158275} {\bibfield  {journal} {\bibinfo  {journal}
  {Science}\ }\textbf {\bibinfo {volume} {321}},\ \bibinfo {pages} {544}
  (\bibinfo {year} {2008})}\BibitemShut {NoStop}%
\bibitem [{\citenamefont {Howell}\ \emph {et~al.}(2004)\citenamefont {Howell},
  \citenamefont {Bennink}, \citenamefont {Bentley},\ and\ \citenamefont
  {Boyd}}]{Howell2004}%
  \BibitemOpen
  \bibfield  {author} {\bibinfo {author} {\bibfnamefont {J.~C.}\ \bibnamefont
  {Howell}}, \bibinfo {author} {\bibfnamefont {R.~S.}\ \bibnamefont {Bennink}},
  \bibinfo {author} {\bibfnamefont {S.~J.}\ \bibnamefont {Bentley}}, \ and\
  \bibinfo {author} {\bibfnamefont {R.~W.}\ \bibnamefont {Boyd}},\ }\href
  {\doibase 10.1103/PhysRevLett.92.210403} {\bibfield  {journal} {\bibinfo
  {journal} {Phys. Rev. Lett.}\ }\textbf {\bibinfo {volume} {92}},\ \bibinfo
  {pages} {210403} (\bibinfo {year} {2004})}\BibitemShut {NoStop}%
\bibitem [{\citenamefont {Leach}\ \emph {et~al.}(2012)\citenamefont {Leach},
  \citenamefont {Warburton}, \citenamefont {Ireland}, \citenamefont {Izdebski},
  \citenamefont {Barnett}, \citenamefont {Yao}, \citenamefont {Buller},\ and\
  \citenamefont {Padgett}}]{Leach2012}%
  \BibitemOpen
  \bibfield  {author} {\bibinfo {author} {\bibfnamefont {J.}~\bibnamefont
  {Leach}}, \bibinfo {author} {\bibfnamefont {R.~E.}\ \bibnamefont
  {Warburton}}, \bibinfo {author} {\bibfnamefont {D.~G.}\ \bibnamefont
  {Ireland}}, \bibinfo {author} {\bibfnamefont {F.}~\bibnamefont {Izdebski}},
  \bibinfo {author} {\bibfnamefont {S.~M.}\ \bibnamefont {Barnett}}, \bibinfo
  {author} {\bibfnamefont {A.~M.}\ \bibnamefont {Yao}}, \bibinfo {author}
  {\bibfnamefont {G.~S.}\ \bibnamefont {Buller}}, \ and\ \bibinfo {author}
  {\bibfnamefont {M.~J.}\ \bibnamefont {Padgett}},\ }\href {\doibase
  10.1103/PhysRevA.85.013827} {\bibfield  {journal} {\bibinfo  {journal} {Phys.
  Rev. A}\ }\textbf {\bibinfo {volume} {85}},\ \bibinfo {pages} {013827}
  (\bibinfo {year} {2012})}\BibitemShut {NoStop}%
\bibitem [{\citenamefont {Lantz}\ \emph {et~al.}(2008)\citenamefont {Lantz},
  \citenamefont {Blanchet}, \citenamefont {Furfaro},\ and\ \citenamefont
  {Devaux}}]{Lantz2008}%
  \BibitemOpen
  \bibfield  {author} {\bibinfo {author} {\bibfnamefont {E.}~\bibnamefont
  {Lantz}}, \bibinfo {author} {\bibfnamefont {J.-L.}\ \bibnamefont {Blanchet}},
  \bibinfo {author} {\bibfnamefont {L.}~\bibnamefont {Furfaro}}, \ and\
  \bibinfo {author} {\bibfnamefont {F.}~\bibnamefont {Devaux}},\ }\href
  {\doibase {10.1111/j.1365-2966.2008.13200.x}} {\bibfield  {journal} {\bibinfo
   {journal} {Monthly Notices Of The Royal Astronomical Society}\ }\textbf
  {\bibinfo {volume} {386}},\ \bibinfo {pages} {2262} (\bibinfo {year}
  {2008})}\BibitemShut {NoStop}%
\bibitem [{\citenamefont {Jedrkiewicz}\ \emph {et~al.}(2012)\citenamefont
  {Jedrkiewicz}, \citenamefont {Blanchet}, \citenamefont {Lantz},\ and\
  \citenamefont {Trapani}}]{Jedrkiewicz2012}%
  \BibitemOpen
  \bibfield  {author} {\bibinfo {author} {\bibfnamefont {O.}~\bibnamefont
  {Jedrkiewicz}}, \bibinfo {author} {\bibfnamefont {J.-L.}\ \bibnamefont
  {Blanchet}}, \bibinfo {author} {\bibfnamefont {E.}~\bibnamefont {Lantz}}, \
  and\ \bibinfo {author} {\bibfnamefont {P.~D.}\ \bibnamefont {Trapani}},\
  }\href {\doibase 10.1016/j.optcom.2011.09.035} {\bibfield  {journal}
  {\bibinfo  {journal} {Optics Communications}\ }\textbf {\bibinfo {volume}
  {285}},\ \bibinfo {pages} {218 } (\bibinfo {year} {2012})}\BibitemShut
  {NoStop}%
\bibitem [{\citenamefont {Blanchet}\ \emph {et~al.}(2008)\citenamefont
  {Blanchet}, \citenamefont {Devaux}, \citenamefont {Furfaro},\ and\
  \citenamefont {Lantz}}]{Blanchet2008}%
  \BibitemOpen
  \bibfield  {author} {\bibinfo {author} {\bibfnamefont {J.-L.}\ \bibnamefont
  {Blanchet}}, \bibinfo {author} {\bibfnamefont {F.}~\bibnamefont {Devaux}},
  \bibinfo {author} {\bibfnamefont {L.}~\bibnamefont {Furfaro}}, \ and\
  \bibinfo {author} {\bibfnamefont {E.}~\bibnamefont {Lantz}},\ }\href
  {\doibase 10.1103/PhysRevLett.101.233604} {\bibfield  {journal} {\bibinfo
  {journal} {Phys. Rev. Lett.}\ }\textbf {\bibinfo {volume} {101}},\ \bibinfo
  {pages} {233604} (\bibinfo {year} {2008})}\BibitemShut {NoStop}%
\bibitem [{\citenamefont {Blanchet}\ \emph {et~al.}(2010)\citenamefont
  {Blanchet}, \citenamefont {Devaux}, \citenamefont {Furfaro},\ and\
  \citenamefont {Lantz}}]{Blanchet2010}%
  \BibitemOpen
  \bibfield  {author} {\bibinfo {author} {\bibfnamefont {J.-L.}\ \bibnamefont
  {Blanchet}}, \bibinfo {author} {\bibfnamefont {F.}~\bibnamefont {Devaux}},
  \bibinfo {author} {\bibfnamefont {L.}~\bibnamefont {Furfaro}}, \ and\
  \bibinfo {author} {\bibfnamefont {E.}~\bibnamefont {Lantz}},\ }\href
  {\doibase 10.1103/PhysRevA.81.043825} {\bibfield  {journal} {\bibinfo
  {journal} {Phys. Rev. A}\ }\textbf {\bibinfo {volume} {81}},\ \bibinfo
  {pages} {043825} (\bibinfo {year} {2010})}\BibitemShut {NoStop}%
\bibitem [{\citenamefont {Devaux}\ \emph {et~al.}(2012)\citenamefont {Devaux},
  \citenamefont {Mougin-Sisini}, \citenamefont {Moreau},\ and\ \citenamefont
  {Lantz}}]{Devaux2012}%
  \BibitemOpen
  \bibfield  {author} {\bibinfo {author} {\bibfnamefont {F.}~\bibnamefont
  {Devaux}}, \bibinfo {author} {\bibfnamefont {J.}~\bibnamefont
  {Mougin-Sisini}}, \bibinfo {author} {\bibfnamefont {P.-A.}\ \bibnamefont
  {Moreau}}, \ and\ \bibinfo {author} {\bibfnamefont {E.}~\bibnamefont
  {Lantz}},\ }\href@noop {} {\bibfield  {journal} {\bibinfo  {journal}
  {(to appear in EPJD).}\ } (\bibinfo {year} {2012})}\BibitemShut {NoStop}%
\bibitem [{\citenamefont {Jost}\ \emph {et~al.}(1998)\citenamefont {Jost},
  \citenamefont {Sergienko}, \citenamefont {Abouraddy}, \citenamefont {Saleh},\
  and\ \citenamefont {Teich}}]{Jost1998}%
  \BibitemOpen
  \bibfield  {author} {\bibinfo {author} {\bibfnamefont {B.}~\bibnamefont
  {Jost}}, \bibinfo {author} {\bibfnamefont {A.}~\bibnamefont {Sergienko}},
  \bibinfo {author} {\bibfnamefont {A.}~\bibnamefont {Abouraddy}}, \bibinfo
  {author} {\bibfnamefont {B.}~\bibnamefont {Saleh}}, \ and\ \bibinfo {author}
  {\bibfnamefont {M.}~\bibnamefont {Teich}},\ }\href {\doibase
  10.1364/OE.3.000081} {\bibfield  {journal} {\bibinfo  {journal} {Opt.
  Express}\ }\textbf {\bibinfo {volume} {3}},\ \bibinfo {pages} {81} (\bibinfo
  {year} {1998})}\BibitemShut {NoStop}%
\bibitem [{\citenamefont {Oemrawsingh}\ \emph {et~al.}(2002)\citenamefont
  {Oemrawsingh}, \citenamefont {van Drunen}, \citenamefont {Eliel},\ and\
  \citenamefont {Woerdman}}]{Oemrawsingh2002}%
  \BibitemOpen
  \bibfield  {author} {\bibinfo {author} {\bibfnamefont {S.~S.~R.}\
  \bibnamefont {Oemrawsingh}}, \bibinfo {author} {\bibfnamefont {W.~J.}\
  \bibnamefont {van Drunen}}, \bibinfo {author} {\bibfnamefont {E.~R.}\
  \bibnamefont {Eliel}}, \ and\ \bibinfo {author} {\bibfnamefont {J.~P.}\
  \bibnamefont {Woerdman}},\ }\href {\doibase 10.1364/JOSAB.19.002391}
  {\bibfield  {journal} {\bibinfo  {journal} {J. Opt. Soc. Am. B}\ }\textbf
  {\bibinfo {volume} {19}},\ \bibinfo {pages} {2391} (\bibinfo {year}
  {2002})}\BibitemShut {NoStop}%
\bibitem [{\citenamefont {van Exter}\ \emph {et~al.}(2006)\citenamefont {van
  Exter}, \citenamefont {Aiello}, \citenamefont {Oemrawsingh}, \citenamefont
  {Nienhuis},\ and\ \citenamefont {Woerdman}}]{Exter2006}%
  \BibitemOpen
  \bibfield  {author} {\bibinfo {author} {\bibfnamefont {M.~P.}\ \bibnamefont
  {van Exter}}, \bibinfo {author} {\bibfnamefont {A.}~\bibnamefont {Aiello}},
  \bibinfo {author} {\bibfnamefont {S.~S.~R.}\ \bibnamefont {Oemrawsingh}},
  \bibinfo {author} {\bibfnamefont {G.}~\bibnamefont {Nienhuis}}, \ and\
  \bibinfo {author} {\bibfnamefont {J.~P.}\ \bibnamefont {Woerdman}},\ }\href
  {\doibase 10.1103/PhysRevA.74.012309} {\bibfield  {journal} {\bibinfo
  {journal} {Phys. Rev. A}\ }\textbf {\bibinfo {volume} {74}},\ \bibinfo
  {pages} {012309} (\bibinfo {year} {2006})}\BibitemShut {NoStop}%
\bibitem [{\citenamefont {Brambilla}\ \emph {et~al.}(2004)\citenamefont
  {Brambilla}, \citenamefont {Gatti}, \citenamefont {Bache},\ and\
  \citenamefont {Lugiato}}]{Brambilla2004}%
  \BibitemOpen
  \bibfield  {author} {\bibinfo {author} {\bibfnamefont {E.}~\bibnamefont
  {Brambilla}}, \bibinfo {author} {\bibfnamefont {A.}~\bibnamefont {Gatti}},
  \bibinfo {author} {\bibfnamefont {M.}~\bibnamefont {Bache}}, \ and\ \bibinfo
  {author} {\bibfnamefont {L.~A.}\ \bibnamefont {Lugiato}},\ }\href {\doibase
  10.1103/PhysRevA.69.023802} {\bibfield  {journal} {\bibinfo  {journal} {Phys.
  Rev. A}\ }\textbf {\bibinfo {volume} {69}},\ \bibinfo {pages} {023802}
  (\bibinfo {year} {2004})}\BibitemShut {NoStop}%
\bibitem [{\citenamefont {Lantz}\ and\ \citenamefont
  {Devaux}(2000)}]{Lantz2000}%
  \BibitemOpen
  \bibfield  {author} {\bibinfo {author} {\bibfnamefont {E.}~\bibnamefont
  {Lantz}}\ and\ \bibinfo {author} {\bibfnamefont {F.}~\bibnamefont {Devaux}},\
  }\href@noop {} {\bibfield  {journal} {\bibinfo  {journal} {J. Opt. A}\
  }\textbf {\bibinfo {volume} {2}},\ \bibinfo {pages} {362} (\bibinfo {year}
  {2000})}\BibitemShut {NoStop}%
\bibitem [{\citenamefont {Brambilla}\ \emph {et~al.}(2008)\citenamefont
  {Brambilla}, \citenamefont {Caspani}, \citenamefont {Jedrkiewicz},
  \citenamefont {Lugiato},\ and\ \citenamefont {Gatti}}]{Brambilla2008}%
  \BibitemOpen
  \bibfield  {author} {\bibinfo {author} {\bibfnamefont {E.}~\bibnamefont
  {Brambilla}}, \bibinfo {author} {\bibfnamefont {L.}~\bibnamefont {Caspani}},
  \bibinfo {author} {\bibfnamefont {O.}~\bibnamefont {Jedrkiewicz}}, \bibinfo
  {author} {\bibfnamefont {L.~A.}\ \bibnamefont {Lugiato}}, \ and\ \bibinfo
  {author} {\bibfnamefont {A.}~\bibnamefont {Gatti}},\ }\href {\doibase
  10.1103/PhysRevA.77.053807} {\bibfield  {journal} {\bibinfo  {journal} {Phys.
  Rev. A}\ }\textbf {\bibinfo {volume} {77}},\ \bibinfo {pages} {053807}
  (\bibinfo {year} {2008})}\BibitemShut {NoStop}%
\end{thebibliography}
\end{document}